\documentclass[10pt,letterpaper]{article}
\usepackage[top=0.85in,left=2.75in,footskip=0.75in]{geometry}
\usepackage{graphicx,caption,subcaption,float}

\usepackage{cite}
\usepackage{nameref,hyperref}

\begin{document}
	\vspace*{0.2in}
	
	\begin{flushleft}
		{\Large
			\textbf\newline{Complexity Matching and Requisite Variety} 
		}
		\newline
		\\
		Korosh Mahmoodi \textsuperscript{1,*},
		Bruce J. West\textsuperscript{2},
		Paolo Grigolini\textsuperscript{3}	
		\\
		\bigskip
		\textbf{1} Department of Social and Decision Sciences, Carnegie Mellon
		University, Pittsburgh, PA 15213, USA\\
	
		\textbf{2} Information Science
		Directorate,  Army Research Office, Research Triangle Park,  NC 27708, USA\\

\textbf{3} Center for Nonlinear Science, University of North Texas, P.O. Box 311427,
Denton, TX 76203, USA\\

\bigskip

* corresponding author, koroshm@andrew.cmu.edu
\end{flushleft}

\section*{Abstract}
Complexity matching characterizes the role of information in interactions between systems and can be traced back to the 1957 Introduction to Cybernetics by Ross Ashby.
We argue that complexity can be expressed in
terms of crucial events, which are generated by the processes of spontaneous
self-organization. Complex processes, ranging from biological to
sociological, must satisfy the homeodynamic condition and host crucial
events that have recently been shown to drive the information transport
between complex systems. We adopt a phenomenological approach, based on the
subordination to periodicity that makes it possible to combine homeodynamics
and self-organization induced crucial events. The complexity of crucial
events is defined by the waiting- time probability density function (PDF) of
the intervals between consecutive crucial events, which have an inverse
power law (IPL) PDF $\psi (\tau )\propto 1/(\tau )^{\mu }$ with $1<\mu <3$. We show that the action of crucial events has an effect compatible with the shared notion of 
complexity-induced entropy reduction, while making the synchronization between systems sharing the same complexity different from chaos synchronization. 
We establish the coupling between two temporally complex systems using a
phenomenological approach inspired by models of swarm cognition and prove
that complexity matching, namely sharing the same IPL index $\mu $, 
facilitates the transport of information, generating perfect
synchronization.  This new form of complexity matching is expected to contribute significantly to progress in understanding and improving biofeedback therapies.\\

\section*{Author summary}
This paper is devoted to the control of complex dynamical systems, inspired by real processes of biological and sociological interest. The concept of complexity we adopt focuses on the assumption that the processes of self-organization generate intermittent fluctuations and that the time interval between consecutive fluctuations is described by an IPL PDF making the second moment of these time intervals diverge. These fluctuations are identified as crucial events and are responsible for the ergodicity breaking that is widely revealed by the experimental observation of biological dynamics.  We argue that the information transport from one to another complex system is ruled by these crucial events and  propose an efficient theoretical prescription leading to qualitative agreement with experimental results, shedding light into the processes of social learning. The theory developed herein is expected to have important medical applications, such as improving biofeedback techniques,  heart-brain  communication and a significant  contribution to understanding how emotions balance cognition.

\section{INTRODUCTION}

The mathematician Norbert Wiener observed \cite{wiener48a} in a 1948
`popular' lecture that the complex networks in the social and life sciences
appear to behave differently from, but not in contradiction to, the laws in
the physical sciences. He makes the point that the force laws and therefore
control of social phenomena do not necessarily follow from changes in
energy, but rather can be dominated by changes in entropy (information). We
refer to this honorifically as the Wiener Rule (WR), since it was presented in the
form of a statement and left unproven. That lecture appeared the same year
he introduced the new science of \textit{Cybernetics} \cite{wiener48b} to
the scientific community in which his interest in control and communication
within and between animals and machines was made clear.

The economic webs of global finance and stock markets; the social meshes of
governments and terrorist organizations; the transportation networks of
planes and highways; the ecowebs of food networks and species diversity; the
physical wicker of the Internet; the bionet of gene regulation, and so on,
have all been modeled using the nascent science of networks. As these
networks in which we are immersed become increasingly complex a number of
apparently universal properties begin to emerge. One of these properties is
a version of the WR having to do with how efficiently interacting complex
networks exchange information with one another.

West et al. \cite{west08}, among many others, noted that complexity often arises in a network
when the power spectrum $S_{p}(f)$ takes on an IPL
shape: 
\begin{equation}
S_{p}(f)\propto \frac{1}{f^{\alpha }},  \label{spectrum}
\end{equation}%
with the IPL index $\alpha $ in the interval $0.5\leq \alpha \leq 1.5$. In
fact, this $1/f-$ variability is taken by many scientists to be the
signature of complexity and appears in a vast array of phenomena including
the human brain \cite{kello07}, body movements \cite{delignieres16}, music 
\cite{pease18,su07,voss75}, physiology \cite{west06}, genomics \cite{li05},
and sociology \cite{sumpter08}. They \cite{west08} then used the IPL index
as a measure of a network's complexity and reviewed the literature arguing
that two complex interacting networks exchange information most efficiently
when the IPL indices of the two networks match. The hypothesis of the 
\textit{complexity management} \textit{effect} form of the WR was proven
by Aquino et al. \cite{aquino10,aquino11} using averages over PDFs and more generally using time averages \cite{piccinini16}.

In less than a decade after the initial formulation of cybernetics, Ross
Ashby, captured in his introduction to the subject \cite{ashby57}, the
difficulty of regulating biological systems and that \textquotedblleft the
main cause of difficulty is the variety in the disturbances that must be
regulated against". This insightful observation, which was subsequently
extended to complex networks in general, led to the conclusion that it is
possible to regulate them if the regulators share the same requisite variety
(complexity) as the systems being regulated. Herein we refer to Ashby's 
\textit{requisite variety} with the more recent term  \textit{complexity matching} \textit{effect} (CME) \cite{west08}. The term complexity matching has been widely used in the recent past to
include such information exchange activities as side-by-side walking \cite
{almurad17}, ergometer rowing \cite{hartigh18}, syncopated finger tapping 
\cite{coey16}, dyadic conversation \cite{abney14}, and interpersonel
coordination \cite{marmelat12,fine15}. These synchronizations are today's
realizations of the regulation of the brain, in conformity with the
observations of Ashby.

\paragraph{Crucial events and the brain:}

\qquad For the purposes of the present paper it is important to stress that
there exists further research directed toward the foundations of social
learning \cite{fonseca16,froese14,dumas10,jaegher10} that is even more
closely connected to Ashby's challenge of the regulator and the regulated
sharing a common level of complexity. In fact, this social learning research
aims at evaluating the transfer of information from the brain of one player
to that of another by way of the interaction the two players established
through their avatars \cite{froese14}, which is to say, it is a virtual
reality experiment capturing the social cognition shared by a pair of
humans. The results are exciting in that the trajectories of pairs of
players turn out to be significantly synchronized. But even more important
than synchronization is the fact that the trajectories of the two avatars
have a universal structure based on the shared EEGs of the paired human
brains.

Herein we provide a theoretical rational for the universal structure
representing the brains of the pairs of interacting individuals, based on
the CME. More broadly the theory can be adapted to the communication, or
information transfer, between the heart and the brain \cite{brain-heart}
of a single individual. The key to this understanding is the existence of
crucial events. In a system as complex as the human brain \cite{righi16} there is
experimental evidence for the existence of crucial events, which, for our
purposes here, can be interpreted as organization rearrangements, or renewal
failures. The time interval between consecutive crucial events is described
by a waiting-time IPL PDF
\begin{equation}
\psi(\tau)\propto 1/t^\mu  \qquad \ with \qquad 1<\mu <3.
\label{waiting}
\end{equation}%
The crucial events generate ergodicity breaking and are widely studied to
reveal fundamental biological statistical properties \cite{metzler14}.

The transfer of information between interacting systems has been addressed
using different theoretical tools, examples of which include: chaos
syncronization \cite{rosenblum96}, self-organization \cite{schwab14}, and
resonance \cite{pariz18}. However, none of these theoretical approaches have
to date explained the experimental results that exist for the correlation
between the dynamics of two distinct physiological systems \cite
{kitzbichler09}. Herein we relate this correlation to the occurrence of
crucial events, which are responsible for the generation of $1/f-$
variability with an IPL spectrum having an IPL index $3-\mu $ \ and for the
results of a number of psychological experiments including those of Correll 
\cite{correll08}. The experimental data imply that activating cognition has
the effect of making the IPL index $\mu <3$ cross the barrier between the L\'{e}vy and Gauss basins of attraction, namely making \ $\mu >3$ \cite
{grigolini09}. This is in line with Heidegger's phenomenology \cite{dotovexperiment}.

The crossing of a basin's boundary is a manifestation of the significant
effect of violating the linear response condition, according to which a
perturbation should be sufficiently weak that it does not affect a system's dynamic
complexity \cite{allegrini09}. The experimental observation obliged us to go
beyond the linear response theory adopted in earlier works in order to
explain the transfer of information from one complex system to another and
to formally prove the WR. This information transfer was accomplished through
the matching of the IPL index of the crucial events PDF of the regulator
with the IPL index of the crucial events PDF of the system being regulated 
\cite{aquino10,piccinini16}. This is consistent with the general idea of the
CME, with the main limitation being that the perturbation intensity is
sufficiently small that it is possible to observe the influence of the
perturbing system on the perturbed system through ensemble averages, namely
a mean over many realizations \cite{aquino10,aquino11}, or through time
averages, if we know the occurrence time of crucial events \cite{piccinini16}.  But, the complexity matching theory of this paper allows us to establish the transfer of information from one complex network to another at the level of single realizations, for example matching between the movements of two Tango dancers.

Note that the 1/f-variability of the spectrum is a necessary, but not a
sufficient, condition to have maximum information exchange between two
complex networks. This is where the present theory deviates from the early
form of cybernetics. The present theory requires the existence of crucial
events.

\paragraph{Homeodynamics:}

\qquad Another important property of biological processes is homeodynamics 
\cite{lloyd01}, which seems to be in conflict with homeostasis as understood
and advocated by Ashby. Lloyd et al. \cite{lloyd01} invoke the existence of
bifurcation points to explain the transition from homeostasis to
homeodynamics. This transition, moving away from Ashby's emphasis on the
fundamental role of homeostasis, has been studied by Ikegami and Suzuki \cite
{ikegami08} and by Oka et al. \cite{oka15}, who coined the term \textit{
	dynamic homeostasis}. They used Ashby's cybernetics to deepen the concept of
self and to establish if the behavior of the Internet is similar to that of
the human brain.

Turalska et al. \cite{turalska09}, based on the direct use of the dynamics
of two complex networks, studied the case when a small fraction of the units
of the regulated system perceive the mean field of the regulating system. At
criticality the choice made by these units is interpreted as \textit{swarm
	intelligence} \cite{vanni11}, and, in the case of the Decision Making Model
(DMM) adopted in \cite{turalska09} is associated with the index $\mu=1.5$.
Synchronization is observed in \cite{turalska09}\  when both systems are in
the critical condition $\mu_{1}=\mu_{2}=1.5$ and it is destroyed if one system is critical 
and the other is  sub-critical, or viceversa.  This suggests that maximal
synchronization is realized when both systems are at criticality, namely,
they share the same IPL index $\mu $.

The present theory covers the complexity matching between networks with different complexity indices $\mu_{1} \neq \mu_{2}$. Also, the present theory is supplemented by homeodynamics, which had not been considered before. This theory
 should not be confused with the unrelated phenomenon of
chaos synchronization. In fact, the intent of the present approach is to
establish the proper theoretical framework to explain, for instance,
brain-heart communication. Here the heart is considered to be a complex, but
not chaotic, system, in accordance with a growing consensus that the heart
dynamics are not chaotic \cite{glass09}.

\section{METHOD} \label{method}

To address Ashby's challenge \cite{ashby57} we adopt the
 perspective of subordination theory \cite{subordination}. This
theoretical perspective is closely connected to the Continuous Time Random
Walk (CTRW) \cite{ctrw}, which is known to generate anomalous
diffusion. We use this viewpoint to establish an 
approach to explaining the experimental results showing the remarkable
oscillatory synchronization between different areas of the brain \cite{kitzbichler09}.

It has to be stressed that a natural choice may rest on the use of Kuramoto's model \cite{kuramoto1}. In fact,  the model of Kuramoto affords a simple paradigm to explain synchronization of rotators, as explained in the excellent review paper of Ref. \cite{kuramoto2}. We think that this popular model can be properly generalized to replace the adoption of its control parameter with the same self-control of Refs. \cite{korosh,social,Resilience} that is shown to generate criticality. The theory of these papers, called Self-Organized-Temporal Crtiticality (SOTC), makes the control parameter spontaneously evolve towards a condition of fluctuation around a critical value that, interpreted as a fluctuating  temperature may lead to physical effects similar to those of the temperature gradients studied in Ref. \cite{kuramoto3}, which affords again a natural way to explain the synchronization of rotators. 
However, this direction, which is left as a subject of future investigation, would make more difficult for us to explain the important role of crucial events for the realization of synchronization. For this reason in this paper we adopt the subordination to a periodic motion of Ref. \cite{ascolani}, which is a generalization of CTRW \cite{ctrw}, based, indeed, on activating the action of crucial events. 

Hereby we describe subordination to periodic and regular rotation. This theory, supplemented by the intelligence necessary to realize CTRW leads us to the  central algorithmic prescription of Eq.  (\ref{central1}) ) and Eq. (\ref{central2}):  there is no progressive phase shift between the clocks if there is no crucial event and a shift if there is. This is the central idea we use to supplement CTRW with the intelligence realizing the spontaneous synchronization that the Kuramoto models generates with a proper choice of the control parameter \cite{kuramoto2}. 

\begin{figure}[h]
	\begin{center}
		\includegraphics[width= 0.5 \linewidth]{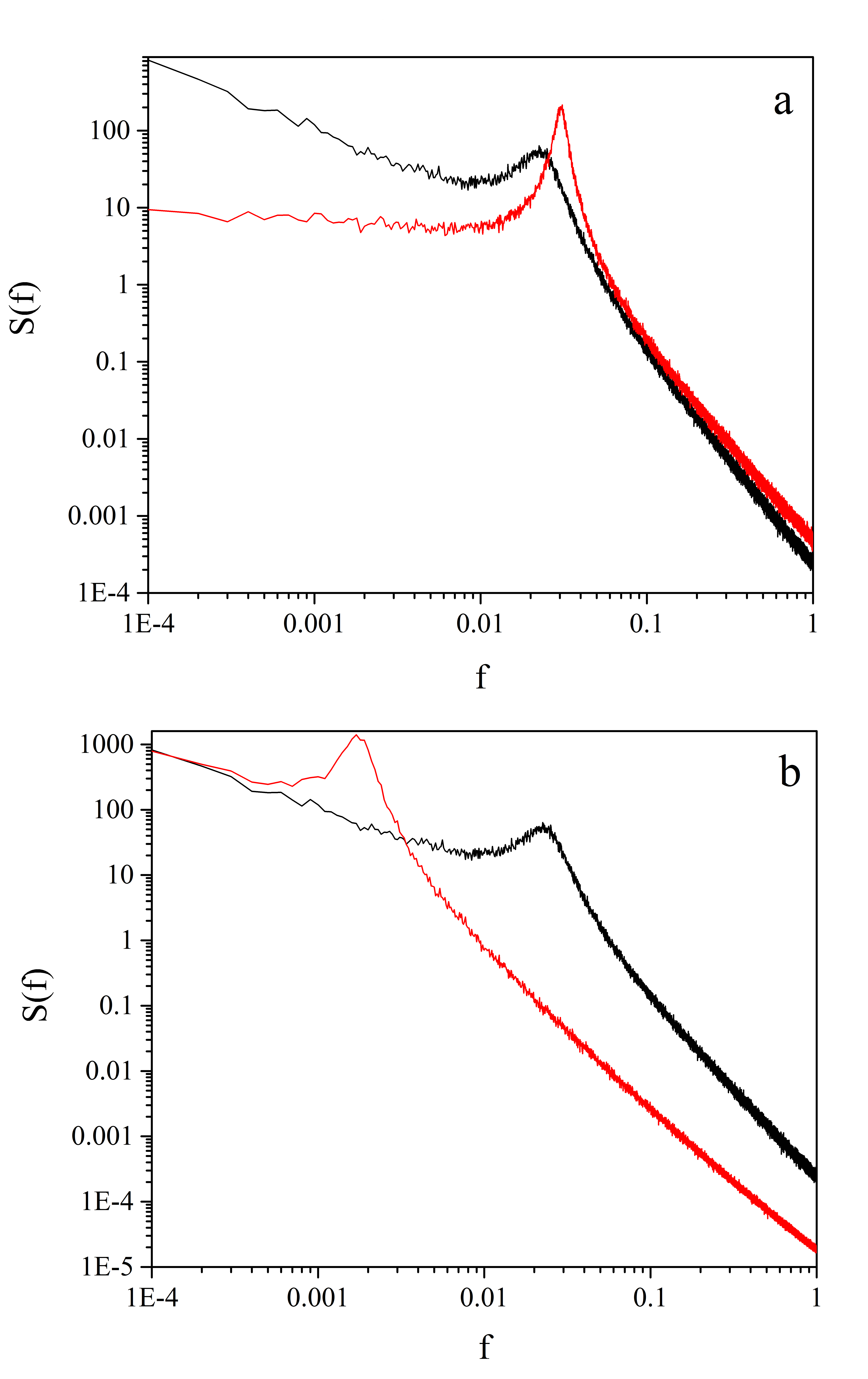}
	\end{center}
	\caption{The spectrum $S(f)$ of 
		subordinations to the regular clock motion. (a) $\Omega =0.06283$, $
		\mu =2.1$ (black curve), $\mu =2.9$ (red curve). (b) $\mu %
		=2.1$, $\Omega =0.06283$ (black curve), $\Omega =0.0006283$ (red curve). }
	\label{UNIVERSAL}
\end{figure}

Consider a clock, whose discrete hand motion is punctuated by ticks and the time interval between
consecutive ticks is, $\Delta t=1$, by assumption. At any tick the angle $%
\theta $ of the clock hand increases by $2\pi /T$, where $T$ is the number
of ticks necessary to make a complete rotation of $2\pi $. We implement
subordination theory by selecting for the time interval between consecutive
ticks a value $\tau / <\tau> $ where $\tau $ is picked from an IPL waiting -time PDF $\psi (\tau )$ with a complexity index $ \mu >2 $. This is a
way of embedding crucial events within the periodic process. Notice that in
the Poisson limit $\mu \rightarrow \infty $ the resulting rotation becomes
virtually indistinguishable from those of the non-subordinated clock. Note
further, that when $\mu >2$, the mean waiting time $\left\langle \tau
\right\rangle $ is finite. As a consequence, if $T$ is the information about
the frequency $\Omega =2\pi /T$, this information is not completely lost in the subordinated time series. 

During the dynamical process the signal frequency fluctuates around $\Omega $
and the average frequency is changed into an effective value 
\begin{equation}
\Omega _{eff}=(\mu -2)\Omega .  \label{david}
\end{equation}
This formula can be easily explained.  In fact, $\mu =3$ is the border between two distinct statistical regions, the Levy and the waiting-time PDF of
the Gaussian region $\mu >3$ where both the first and second moment of $\psi
(\tau )$ are finite, and the average of the fluctuating frequencies is
identical to $\Omega $. In the region $\mu <2$ the process is non-ergodic,
the first moment $<\tau >$ is divergent and the direct indications of
homeodynamics vanish. The condition $2<\mu <3$ is compatible with the
emergence of a stationary correlation function, in the long-time limit, with 
$\mu $ replaced by $\mu -1$. Thus, using the result of earlier work \cite{ascolani} we obtain for the equilibrium correlation function exponentially
damped regular oscillations. At the end of this oscillatory process, an IPL
tail proportional to $1/t^{\mu -1}$is obtained. Using a Tauberian theorem
explains why the power spectrum $S(f)$ becomes proportional to $1/f^{3-\mu }$
for $f\rightarrow 0$. In summary, in a log-log representation, we obtain a
curve with different slopes: $\beta =3-\mu $, to the left of the
frequency-generated bump, and $\beta =2$, to its right. The slope $\beta =2$
is a consequence of the exponentially damped oscillations.

These predictions are confirmed in \nameref {UNIVERSAL}, which illustrates the result of a numerical approach to the subordination to a periodic clock with frequency $\Omega$. 
In the absence of subordination the projection on the abscissa axis would generate $x(t)  = Re (e^{i{\Omega t}})$. When we apply subordination to this regular motion, we interpret the time evolution of $x(t)$ as the result of a cooperative interaction
between many oscillators. The IPL index $\mu $ quantifies the temporal
complexity, spontaneously realized as an effect of oscillator-oscillator
interactions. \nameref {UNIVERSAL} shows that as an effect of periodicity, the low-frequency region where the ideal $1/f$ noise is expected to emerge at $\mu =2$, can be strongly reduced giving more room to the region of noise $1/f^2$ (see panel b).

To make network-1 (S$_{1})$ drive network-2 (S$_{2}$) we have to
generalize the swarm intelligence prescription adopted in earlier work \cite{social, Resilience, korosh, turalska09,vanni11}. This generalization is necessary because
the earlier work was limited to matching of two identical networks at their criticality, and also was based on the assumption that the single units of the
complex networks, in the absence of interaction, undergo dichotomous
fluctuations without the periodicity imposed here. In the absence of
periodicity, the mean field $x(t)$ of the complex network can be written as $x(t)=(U(t)-D(t))/N$, where $U(t)$ is the number of individual in
the state $``Up"$, $x>0$, and $D(t)$ is the number of individuals  
$``Down"$, $x<0$. In the case considered herein the number of units in a network $N=U(t)+D(t)$ is constant. Using this notation (see Section \ref{Supporting})
we show that network-2 under influence of network-1 changes as:
\begin{equation}
\Delta x_2 \propto K(t),  \label{proportionality}
\end{equation}%
where 
\begin{equation} \label{importantprescription}
K(t)\equiv 2(x_1(t) - x_2(t)). 
\label{noperiodicity}
\end{equation}%
To properly take periodicity into account note that the mean field in S$_{1}$
given by $x_{1}(t)$ has the functional form 
\begin{equation}
x_{1}(t)=cos(\Omega _{1}n_{1}(t)).
\end{equation}%
The mean field in S$_{2}$ has the same periodic functional form, up to a
time-dependent phase, 
\begin{equation}
x_{2}(t)=cos(\Omega _{2}n_{2}(t)+\Phi (t)).  \label{phase}
\end{equation}%
The phase $\Phi (t)$ is a consequence of the fact that the units of S$_{2}$
try to compensate for the effects produced by the two independent
self-organization processes. The number of ticks of the S$_{1}$ clock, $%
n_{1}(t)$, due to the occurrence of crucial events, becomes increasingly
different from the number of ticks of the S$_{2}$ clock, $n_{2}(t)$. The
units of S$_{2}$ try to imitate the choices made by the units of S$_{1}$. This is modeled by adjusting 
the phase $\Phi (t)$ of Eq. (\ref{phase}). The phase change is proportional to $K(t)$ and to the derivative of $x_{2}(t)$ with respect to t. Thus, we obtain
the central algorithmic prescription of this paper: 
\begin{equation} \label{central1}
\Phi (t+1)=\Phi (t),
\end{equation}%
if at $t+1$ no crucial event occurs, and 
\begin{equation} \label{central2}
\Phi (t+1)=\Phi (t)-r_{1}K(t)sin\left( \Omega _{2}n_{2}(t)+\Phi (t)\right) ,
\end{equation}%
if at $t+1$ a crucial event occurs. Note that the real positive number $r_{1} < 1
$, defines the proportionality factor left open by Eq. (%
\ref{proportionality}), or, equivalently, defines the strength of the
perturbation that S$_{1}$ exerts on S$_{2}$.

\begin{figure}[h]
	\begin{center}
		\includegraphics[width= 0.7 \linewidth]{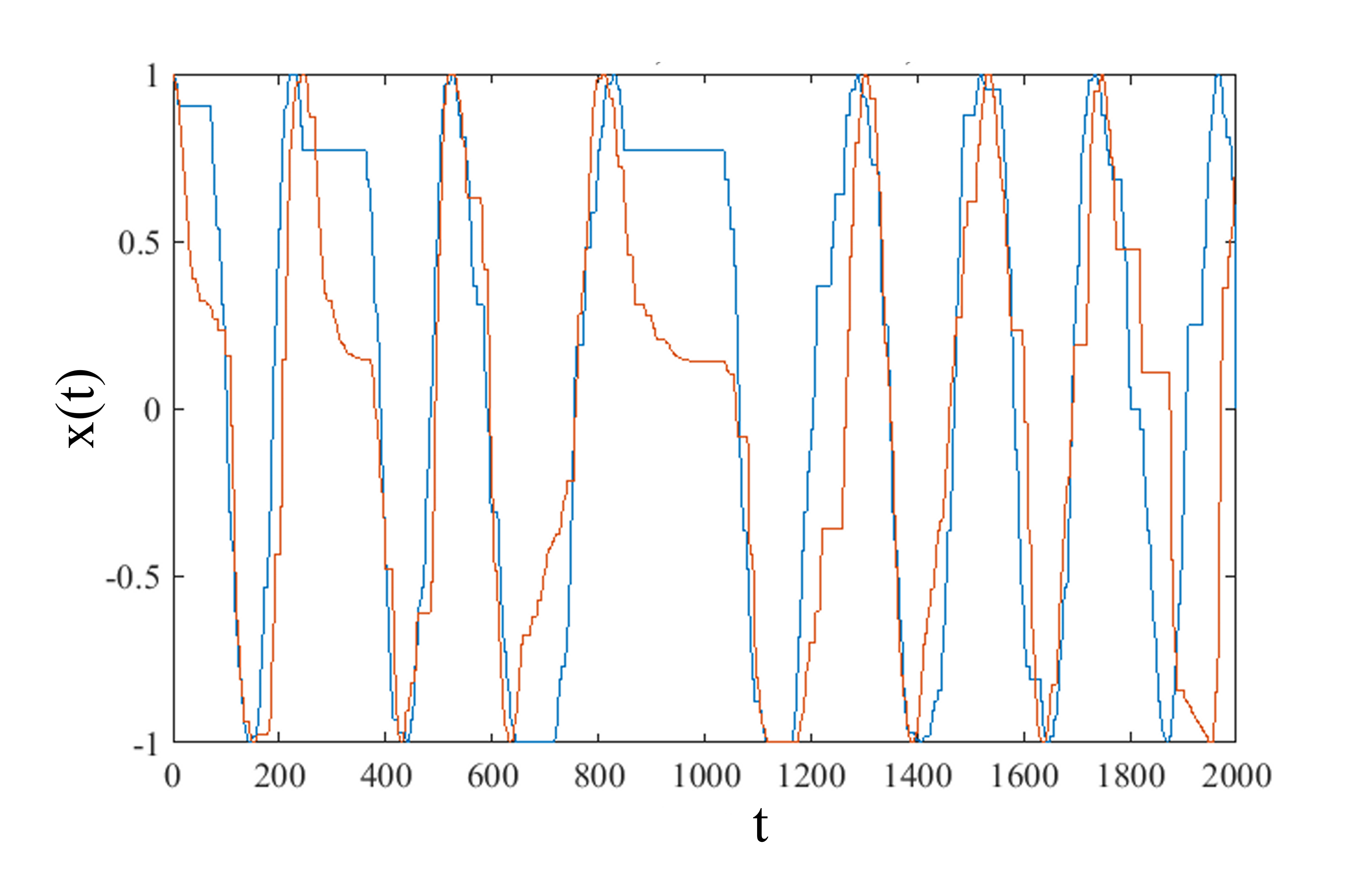}
	\end{center}
	\caption{S$_{1}$ (blue curve) drives S$_{2}$ (red curve). Two
			systems are identical: $\protect\mu =2.2$, $\Omega =0.063$, $r_{1}=0.05$.
			The connection (one directional) is realized using Eq. (\protect\ref{proportionality}) and Eq. (%
			\protect\ref{noperiodicity}). }
	\label{driving}
\end{figure}

 \nameref{driving} illustrates the significant synchronization between
the driven and the driving system obtained for $\mu =2.2$, close to the
values of the crucial events of the brain dynamics \cite{righi16}. This
result also can be used to explain the experimental observation of the
synchronization of two people walking together \cite{almurad17} (see
Section \ref{Supporting}).

\begin{figure}[h]
	\begin{center}
		\includegraphics[width= 0.7 \linewidth]{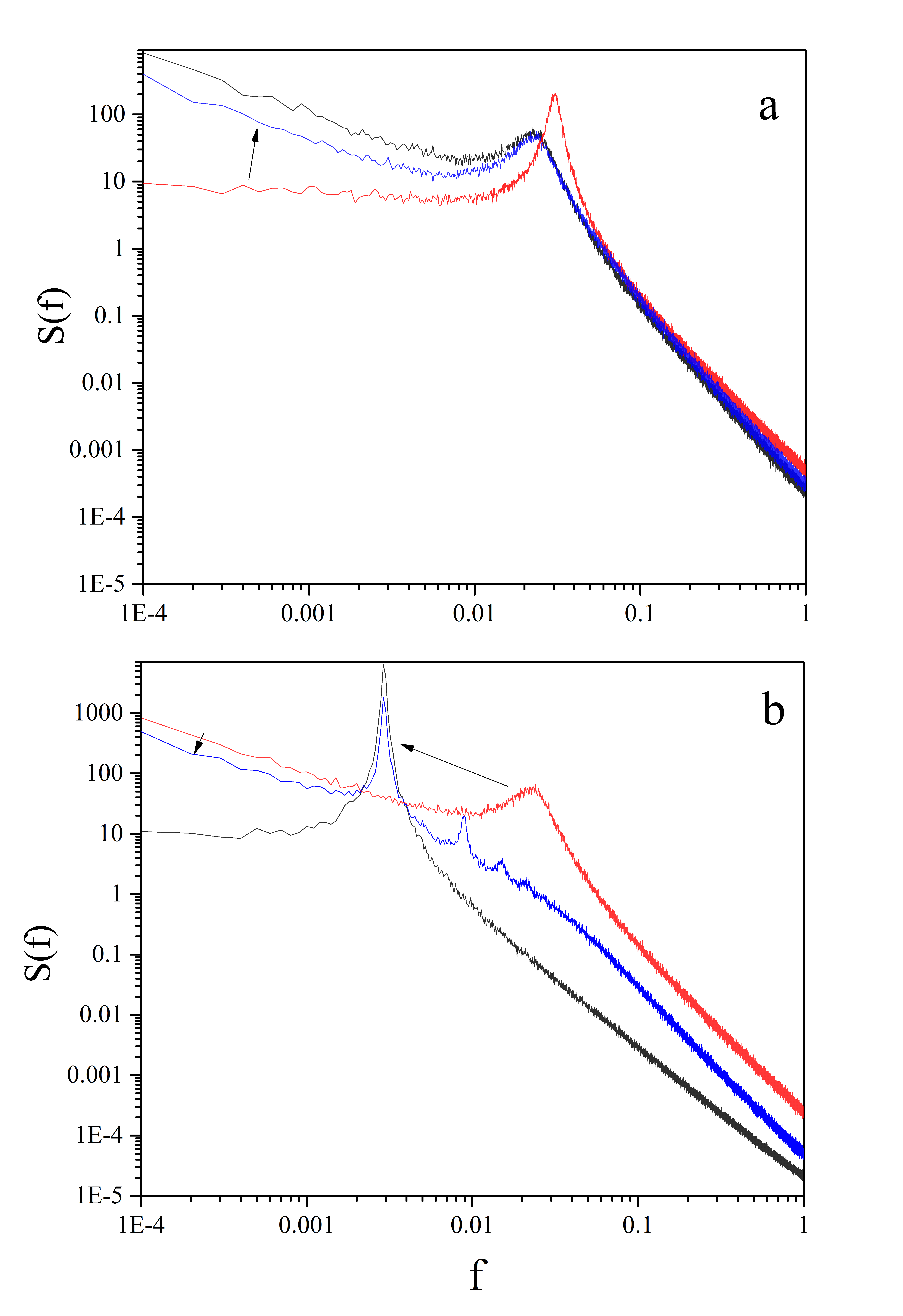}
	\end{center}
	\caption{The spectrums of subordinations. (a) Black curve ($S_{1}$): $\mu =2.1 ,\Omega =0.063$; red curve ($S_{2}$): $\mu 
			=2.9,\Omega =0.063$; blue curve: $S_{2}$ after being connected (one directional) to $S_{1}$ with $
			r_{1}=0.1$. (b) Black curve ($S_{1}$): $\mu =2.9,$ $\Omega =0.0063$; red curve ($S_{2}$): $\mu =2.1,$ $\Omega =0.063$; blue curve: $S_{2}$ after being connected (one directional) to $S_{1}$ with $r_{1}=0.1$.}
	\label{complexdrivingsimple}
\end{figure}

The top panel of  \nameref{complexdrivingsimple} shows that S$_{2}$,
with $\mu _{2}=2.9$, is very close to the Gaussian border and adopts the higher
complexity of S$_{1}$ with $\mu _{1}=2.1$, namely the complexity of a network
very close to the ideal condition, $\mu =2$, to realize $1/f$ noise.

In the bottom panel of  \nameref{complexdrivingsimple} we see that a
driving network very close to the Gaussian border does not make the driven
network less complex, but it does succeed in forcing it to adopt the
regulator's periodicity. Here we have to stress that the perturbing network
is quite different from the external fluctuation that was originally adopted
to mimic the effort generated by a difficult task \cite{correll08,grigolini09}. In
this latter case, according to Heidegger's phenomenology \cite{dotovexperiment} the transition from \emph{ready-to-hand} to \emph{unready-to-hand} makes the IPL index $\mu $ depart from the $1/f$-noise condition $\mu =2$ \cite{correll08,grigolini09} so as to reach the Gaussian border $\mu = 3$ and to go beyond it. Here the perturbation is characterized by an intense periodicity and while it does not change  the complexity of the perturbed network very much, it does transfer its own periodicity.

\begin{figure}[h]
	\begin{center}
		\includegraphics[width= 0.7 \linewidth]{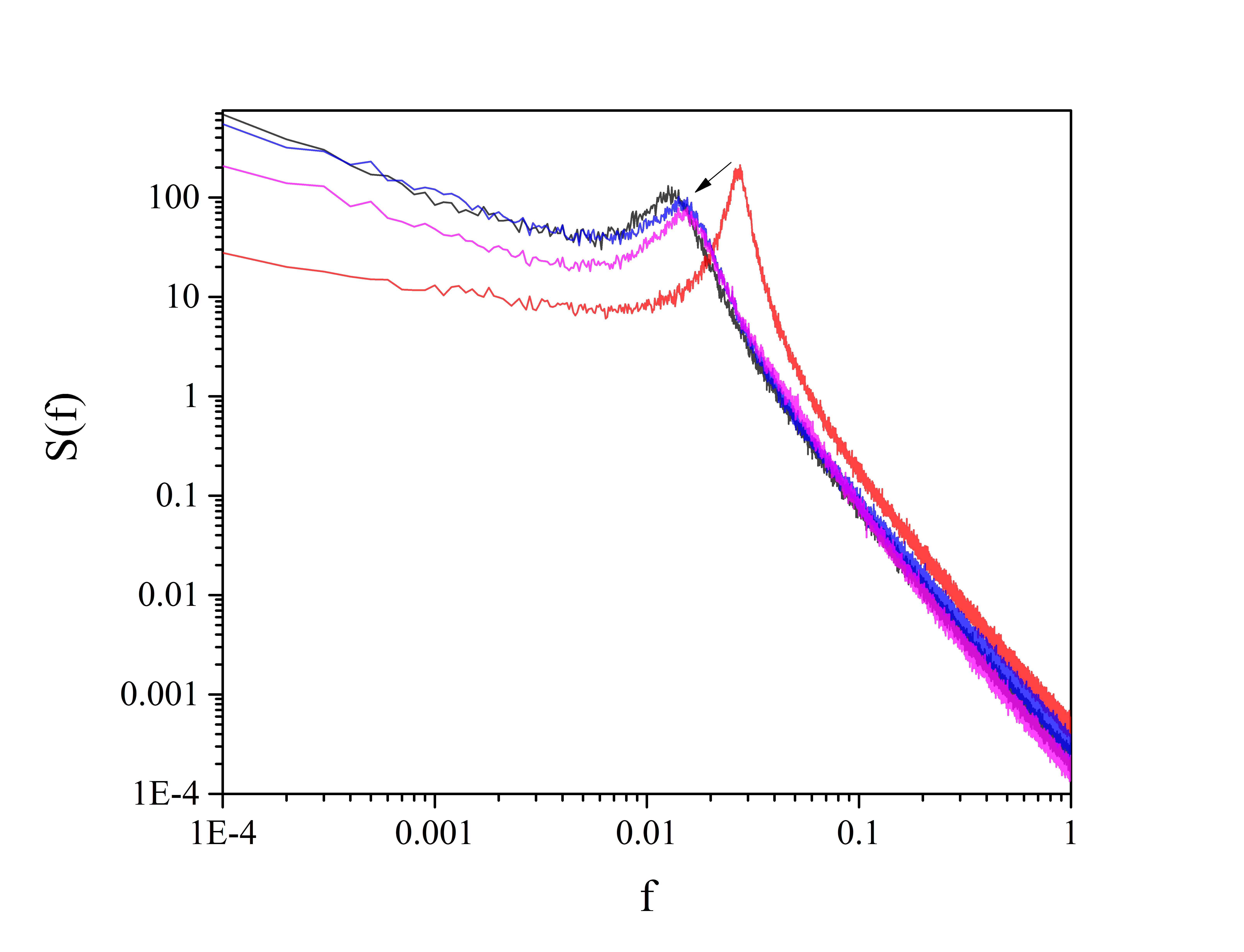}
	\end{center}
	\caption{The spectrums of subordinations. Black curve ($S_{1}$): $\mu _{1}=2.2$, $\Omega _{1}=0.063$; red curve ($S_{2}$): $\mu _{2}=2.9$, $\Omega_{2}=0.063$. Blue and pink curves are the spectrums of $S_{1}$ and $S_{2}$ after being connected (bidirectional) with $r_{1}=r_{2}=0.1$ respectively. }
	\label{resonance2}
\end{figure}

The theory developed herein may shed light on the crucial role of cooperation.  Recent psychological research on collective intelligence \cite{pentland} shows that a cooperative interaction  between the members of a group may improve the global intelligence of a group. To realize a condition that is close to that of Ref. \cite{pentland} we study the case 
where S$_{1}$ is influenced by S$_{2}$ in the same way S$_{2}$ is influenced
by S$_{1}$. To make this extension we have to introduce the new parameter $r_{2}$, which defines the intensity of the influence of S$_{2}$ on S$_{1}.$
As a result of this mutual interaction, we have $\mu _{1}\rightarrow \mu
_{1}^{\prime }$ and $\mu _{2}\rightarrow \mu _{2}^{\prime }$. When $\mu
_{1}<\mu _{2}$ we expect 
\begin{equation}
\mu _{1}<\mu _{1}^{\prime }<\mu _{2}^{\prime }<\mu _{2}.
\end{equation}
 \nameref{resonance2} shows that $\mu _{1}^{\prime }\approx \mu _{1}$,
thereby suggesting that the system with higher
complexity does not perceive its interaction with the other system as a
difficult task, which would force it to increase its own $\mu$ \cite{correll08,grigolini09}, while the less complex system has a sense of relief. We
interpret this result as an important property that should be the subject of
psychological experiments similar to that of Ref. \cite{pentland} to shed light on the mechanisms facilitating the controlled exchange of information in the
teaching and learning processes. The theory underlying \textit{complexity
matching} and therefore requisite variety, makes it possible to go beyond the limitation of the earlier work
on complexity management, as illustrated in Section \ref{Supporting}.

The term \textquotedblleft intelligent" that we are using herein is equivalent
to assessing a network to be as close as possible to the ideal condition $\mu
=2$, corresponding to the ideal $1/f$ noise.  \nameref {resonance2} shows that as a result interaction, the complexity of $S_1$ remains virtually unchanged while that of $S_2$ significantly increases thereby generating a mean value of $\mu$ that is closer to the ideal condition of full intelligence $\mu = 2$. 

 Note that in the same sense two very
intelligent networks are the brain and heart that when healthy share the
property of a $\mu $ being close to $2$. The argument presented herein therefore provides a rationale for (an explanation of) the
synchronization between the heart and brain time series \cite{brain-heart} showing that the
concept of resonance, based on tuning the frequency of the stimulus to that
of the network being perturbed, may not be appropriate for complex biological
networks. Resonance is more appropriate for a physical network, where the
tuning has been adopted over the years for the transport of energy not
information. The widely used therapies resting on biofeedback \cite{chinesebiofeedback},  are  the subject of appraisal \cite{papo} and the present results may contribute 
to making therapeutic progress by establishing their proper use.

\section{Supporting Information} \label{Supporting}

\begin{figure}[h]
	\begin{center}
		\includegraphics[width= 0.7 \linewidth]{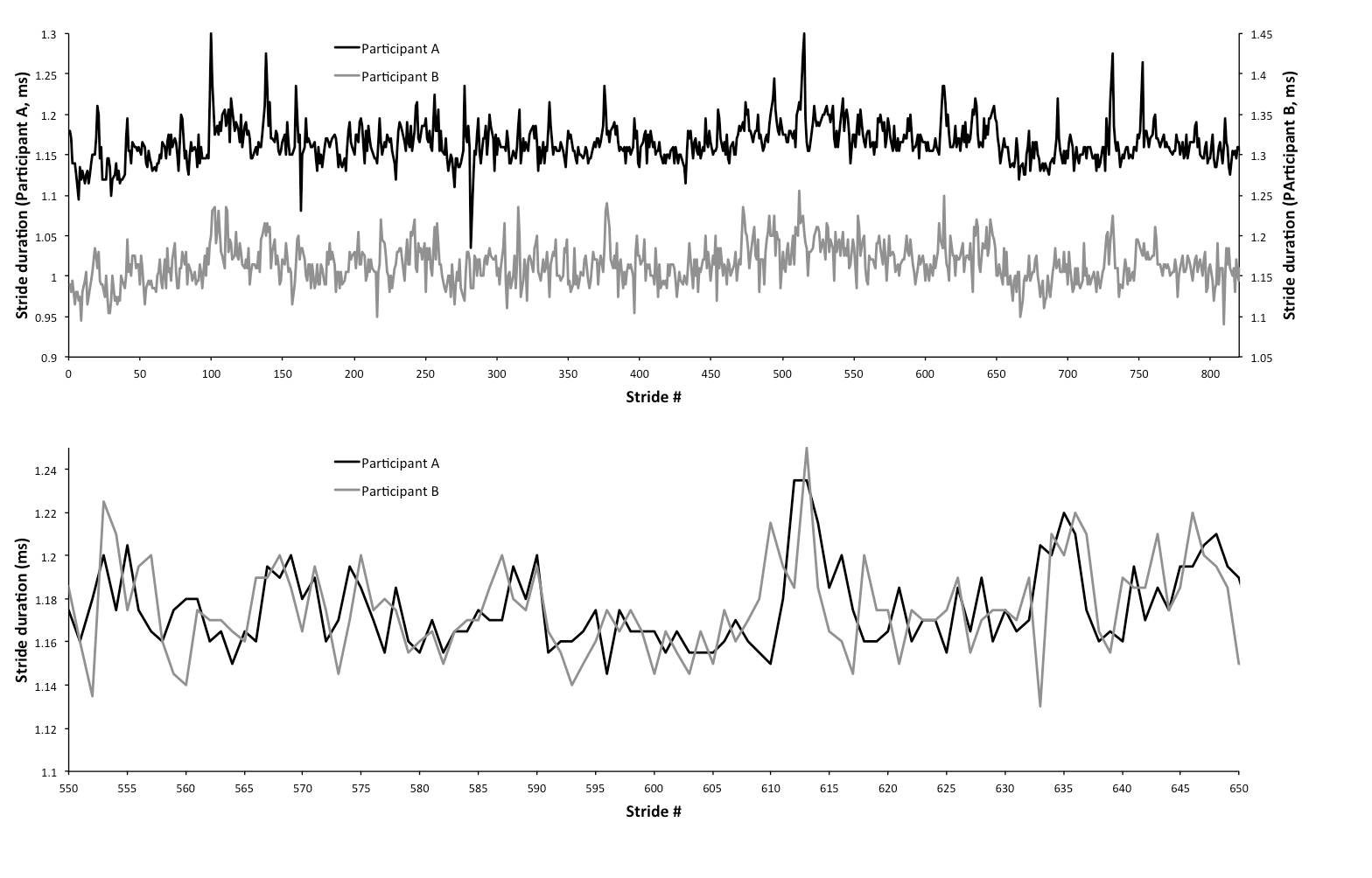}
	\end{center}
	\caption{Experimental walking synchronization. These results have been derived with permission from Ref. \cite{almurad17}. The top panel shows two distinct walking trajectories. These are two human subjects trying to walk together. The bottom panel shows the same trajectories so as to emphasize their synchronization. }
	\label{experimentalwalking}
\end{figure}

This section affords an example of the application of the theory developed herein to the analysis of experimental data.  We focus on the close connection between Fig. 2 of this paper and Fig. 3 of Ref. \cite{almurad17}. For reader's convenience we illustrate this connection with the help of  \nameref{experimentalwalking}. 

\begin{figure}[h]
	\begin{center}
		\includegraphics[width= 0.7 \linewidth]{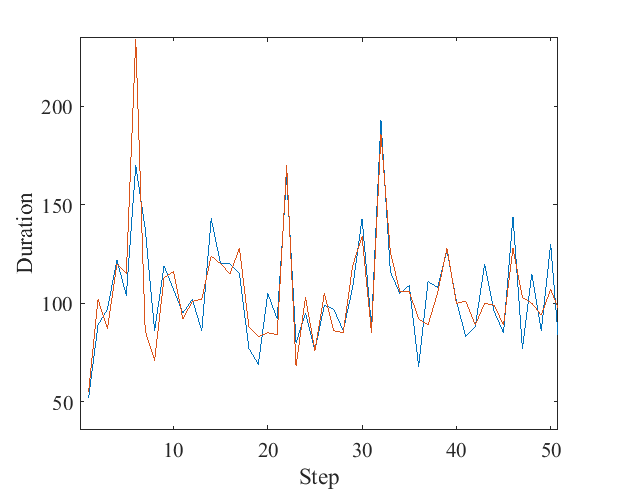}
	\end{center}
	\caption{Time difference between the events of two identical systems connected back to back, $\mu_1 = \mu_2 = 2.2$, $\Omega _{1}= \Omega _{2} = 0.062$, $r_1 = r_2 = 0.1$. }
	\label{theoreticalwalking}
\end{figure}

This figure is the result of the real experiment of Ref. \cite{almurad17} and it should be compared to the qualitatively similar \nameref{theoreticalwalking} obtained with the theory of this paper.

We obtain \nameref{theoreticalwalking} using  Eqs. (2-5) of the text and hereby 
we afford details on how to derive  these important equations. We use  numerical results of the same kind as those illustrated in Fig. 2  properly modified to connect the two trajectories back to back. To make the qualitative similarity with the results of the experiment  \cite{almurad17} more evident we adopt the same prescription as that used by Deligni\'{e}res and his co-worker and interpret the time interval between consecutive crossings of the origin, $x = 0$, of Fig. 2  as the time duration of a stride. We evaluate the mean stride duration and for both the driven and the driving, for any stride we plot the deviation from  mean value.  \nameref{theoreticalwalking} illustrates the result of this procedure, which
 can also be used to explain the synchronization between heart and brain \cite{brain-heart}.

\subsection{Group intelligence}
Although subordination theory does not explicitly depend on the interaction between different units with their own periodicity, S$_{2}$  is driven by S$_{1}$ with  a prescription inspired to create a swarm intelligence \cite{social, Resilience, korosh}. At a given time the units of the driven systems look at the  driving system and according to 
its state increase or decrease the phase of the driven system as described in the Methods section. 

The single individuals of the complex network may have only the value $1$, cooperation,  or $-1$, defection.  We introduce the angle $\theta$ to take periodicity into account and interpret $cos \theta$ as the ratio of the difference between the number of cooperators and the number of defectors to the total number of units. 
Thus  the majority of  cooperators corresponds to $0< \theta < \pi$ and the majority of defectors corresponds  to $\pi < \theta < 2\pi$.
We express that the change in S$_{2}$ because of interaction with S$_{1}$ as
\begin{equation}
\Delta x_2 \propto p(D_2 \rightarrow U_2) - p(U_2\rightarrow D_2),
\label{mines}
\end{equation} 
where the probability of making a transition from the state down to the state up in S$_{2}$ is given by
\begin{equation} \label{plug1}
p(D_2 \rightarrow U_2)  = \frac{D_2}{U_2 + D_2} \frac{U_1}{U_1 + D_1}.
\label{dot}
\end{equation}
The form of Eq. (\ref{dot}) is due to the fact that this probability is the product of the probability of finding a unit in the driven system in the down state by the probability of finding a unit in the driving system in the up state.
Using the same arguments we find
\begin{equation} \label{plug2}
p(U_2 \rightarrow D_2)  = \frac{U_2}{U_2 + D_2} \frac{D_1}{U_1 + D_1}.  
\end{equation}
Let us plug Eq. (\ref{plug1}) and Eq. (\ref{plug2}) into Eq.(\ref{mines}). 
We obtain
\begin{equation}
K(t)\equiv (1-x_{2}(t))(1+x_{1}(t))-(1+x_{2}(t))(1-x_{1}(t)), 
\label{noperiodicity}
\end{equation}
which is the important prescription of Eq. (\ref{importantprescription}).

\subsection{Walking together}
To facilitate appreciation of the similarity between the complexity matching prescription of this paper and the walking synchronization of Ref. \cite{almurad17}, we invite the readers to look at  the experimental results of  \nameref{experimentalwalking}. The real data are not available to us, and we use surrogate data instead. These surrogate data are derived from the numerical results adopted to get  \nameref{theoreticalwalking}, with two major adjustmentsm, as earlier explained, of focusing on the observation of the origin crossing. The remarkably good qualitative agreement between  \nameref{theoreticalwalking} and  \nameref{experimentalwalking} proves the efficiency of the complexity matching approach of this paper.

\subsection{Beyond Complexity Management}

Complexity management is difficult to observe, since it is based on ensemble averages, thereby requiring the average over many identical realizations \cite{aquino11}. In the case of experimental signals of physiological interest, for instance time series relating to brain dynamics, taking the ensemble average is not possible. Recently a procedure was proposed \cite{piccinini16} to convert an individual time series into many independent sequences, so as to again have recourse  to an average over many realizations. This procedure, however, requires  knowledge of time occurrence of crucial events. The theory developed herein makes it possible to evaluate the correlation between the driving and the driven networks using only one realization. We stress that while complexity management \cite{aquino11} does not affect the power index $\mu$ of the interacting complex networks, the present theory, as shown by  \nameref{complexdrivingsimple}, affords important information on how the cooperative interaction makes the unperturbed values of $\mu$ change. 

\begin{figure}[h]
	\begin{center}
		\includegraphics[width= 0.7 \linewidth]{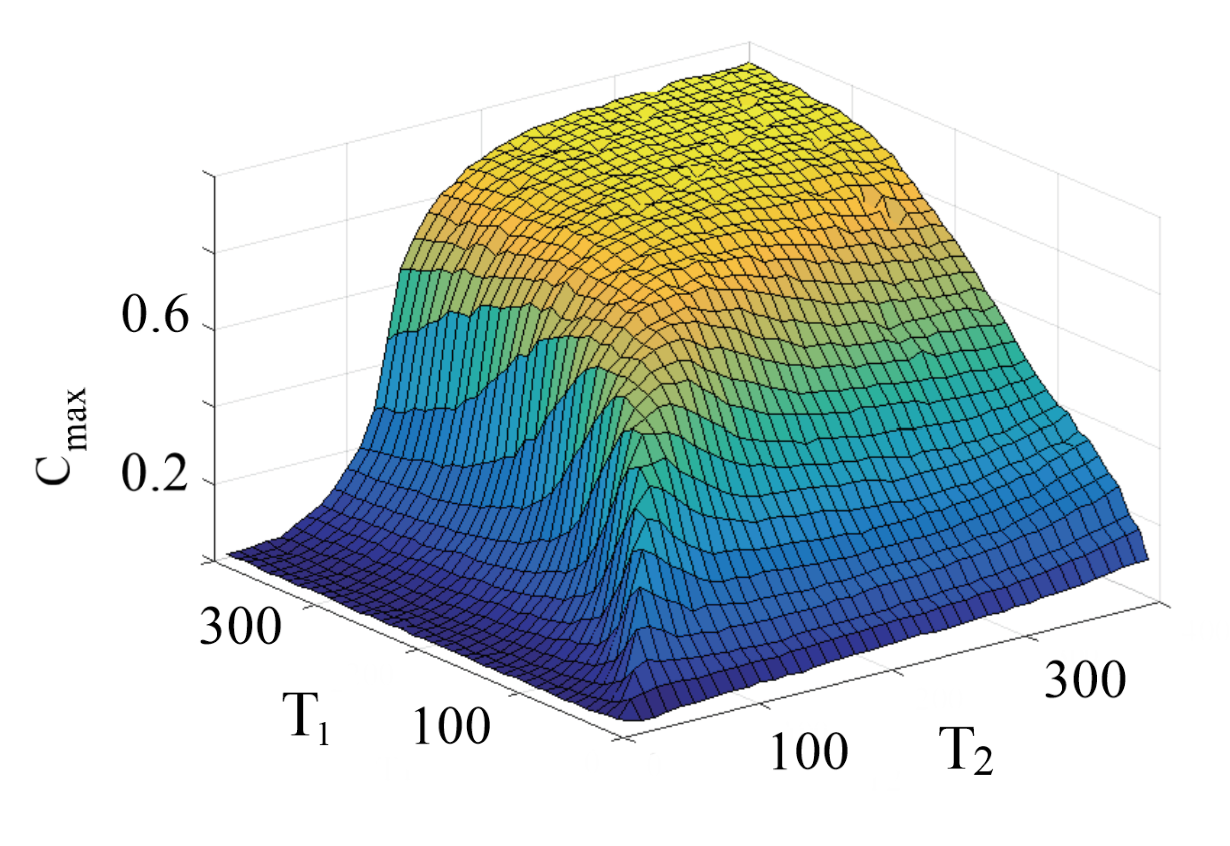}
	\end{center}
	\caption{Dependence of $C_{max}$ (as a measure for complexity matching) on the periodicity of the drive and driven systems. $\mu_{1} = \mu_{2} = 2.8$. $r_1 = 0.1$.}
	\label{resonance}
\end{figure}

In  \nameref{resonance} the maximum value of the cross correlation function, $C_{max}$, between the driving and the driven. This figure   shows that a significantly large frequency mismatch strongly reduces the intensity of synchronization.

\begin{figure}[h]
	\begin{center}
		\includegraphics[width= 0.7 \linewidth]{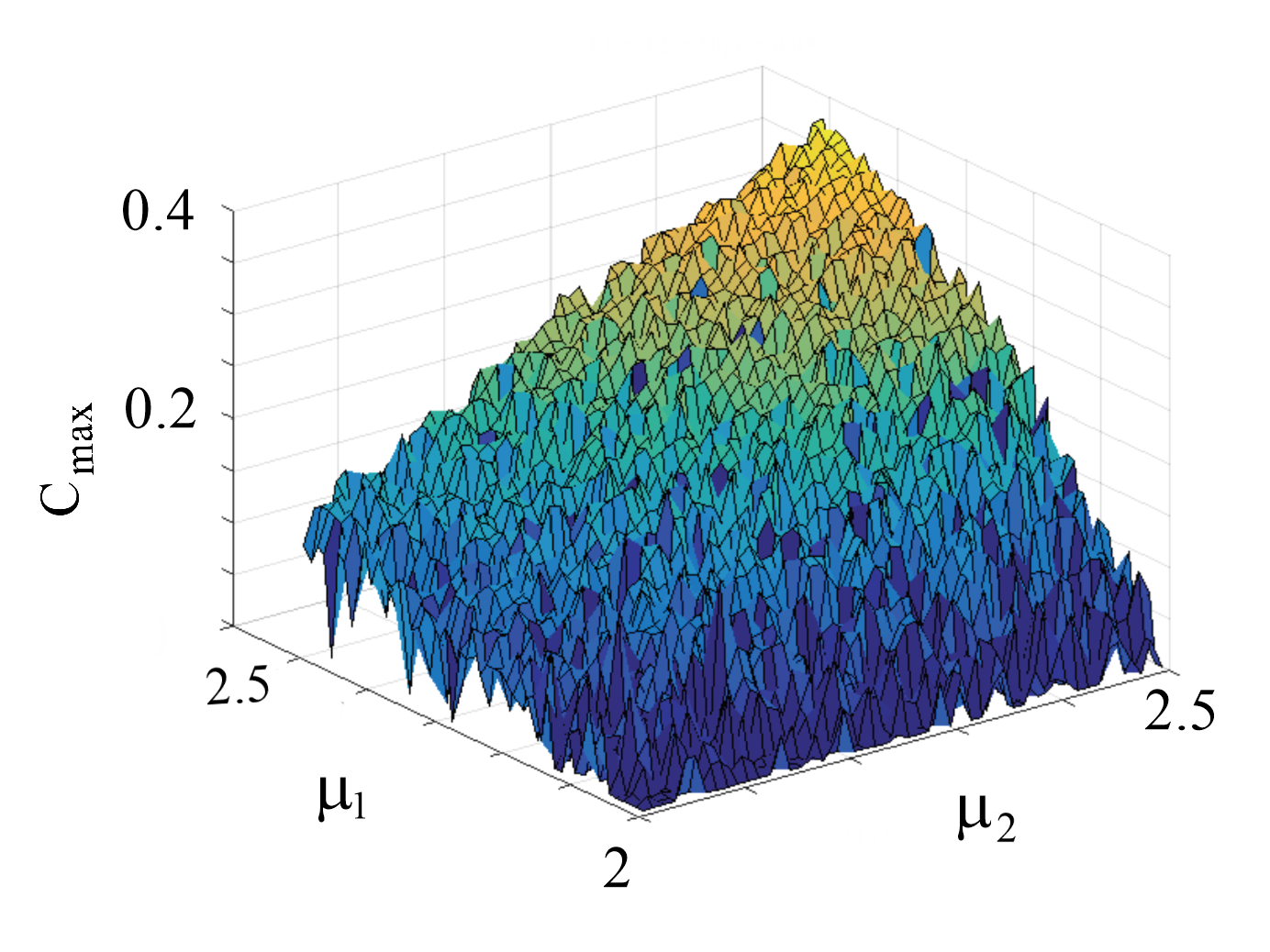}
	\end{center}
	\caption{Dependence of $C_{max}$ on the complexity index of the drive and driven networks. $T_{1}= T_{2} = 50$, $r_1 = 0.1$. }
	\label{changingmu}
\end{figure}

 \nameref{changingmu} shows the effect of changing $\mu_{1}$ and $\mu_{2}$ on $C_{max}$, while keeping the frequencies $\Omega$ identical. 

\section{Complexity, Information and Conclusions}

In the recent literature on self-organization, see, for example,  Gershenson and Fern\'{a}ndez \cite{carlos}, emergence of complexity is interpreted as corresponding to
information reduction, while emphasizing Ashby's concept of homeostasis. Variety  increases with a complex system performing multitask actions and decreases with a complex system focusing on a single task \cite{multitask}. More recent work confirms this property in sociological systems \cite{zhang} while it is well known that it holds true for physiological processes \cite{physiology1,physiology2}. The hypothesis of self-organization has been known and used in biology for nearly half a century \cite{eigen1,eigen2} (see also Chapter 5 of Eigen's important book  \cite{eigen3}). 

The theoretical approach to the complexity matching presented herein may afford a unifying view accounting for the main properties emphasized by this literature while adapting the homeostasis perspective of Ashby \cite{ashby57} to the concept of homeodynamics. Consider the four distinct properties stressed in the current literature: Information reduction, requisite variety, multitasks, homeostasis.

\emph{Information reduction}:
The entropic approach used to deal with  crucial events is the Kolmogorov-Sinai (KS) entropy  $h_{KS}$ \cite{rosa}, which is well described by the formula 
\begin{equation} \label{lyapunov} 
h_{KS}  =  z(2-z) ln2,
\end{equation}
where  $z \equiv \frac{\mu}{\mu-1}$. Eq.(\ref{lyapunov}) indicates that the KS entropy
 vanishes at $z =2$ and it remains equal to $0$ in the whole infinite interval $2 < z < \infty$ ($\mu < 2$). 
Allegrini et al. \cite{compressibility} noticed that $z = 1$, corresponding to $\mu = \infty$,  is the condition of total randomness, namely, the case where an infinitely large amount of information is necessary to control the system. The condition $z = 1.5$, corresponding to $\mu = 3$, makes the sequence of crucial events compressibile, namely, it reduces the amount of information necessary to control the system, and finally 
the $KS$ entropy vanishes when $\mu < 2$.  This is a region characterized  by the diverging value of $<\tau>$. 

The recent generalization to the mechanism of self-organized criticality given by SOTC \cite{korosh} generates crucial events with $\mu < 2$, and, albeit a form of self-organization yielding values of $\mu$ in the interval $2<\mu<3$ is not yet known, we make the plausible conjecture that complex processes that are experimentally proven to generate crucial events in this interval as well as in the interval $1<\mu<2$, are the result of a process of self-organization. The condition $z > 2$ ($\mu < 2)$ is where 
Korabel and Barkai \cite{korabel} had to modify Eq. (\ref{lyapunov}) leaving this expression unchanged for $1 < z < 2$ and making it increase from the  vanishing value  with $z > 2$.  Actually $KS$ entropy is a Lyapunov coefficient and Korabel and Barkai  
defined the Lyapunov coefficient for $z > 2$, by comparing  the departure between two trajectories  moving from very close initial conditions to $t^{\mu-1}$, rather than to  $t$, as correctly done for $z  <  2$. 
This means that the region $z >2$ ($\mu < 2$) is not fully deterministic,  but the amount of information necessary to control the system is drastically reduced. 

\emph{Requisite Variety}: 
Ivanov et al. \cite{nature} noticed that the healthy heartbeats have a variability that makes it impossible to adopt the conventional method of analysis of anomalous scaling based on the stationary assumption.  Consequently they made the assumption of a scaling fluctuation that led them to adopt a multi-fractal approach. Their proposal turned out to be very successful and was adopted to distinguish healthy heartbeats from heart failure heartbeats \cite{nature}. Allegrini et al. \cite{allegrini} examined the same patients studied in \cite{nature} using the crucial events defined herein and found that healthy patients have a $\mu$ very close to the value $\mu= 2$, which makes the  $KS$ entropy vanish. They  also  conjecture that  a self-organization generating crucial events may also be the generator of multi-fractality. This conjecture has been fully confirmed by the recent work of Ref. \cite{Multifractal, bohara}

 Of remarkable importance for the requisite variety issue is the work by Struzik \emph{et al} \cite{physiology2}, emphasizing the transition from $1/f$ noise to $1/f^2$ noise as a manifestation of variability suppression. Healthy heart physiology is based on the balance between the conflicting action of the sympathetic  and parasympathetic nervous systems, thereby resulting in the ideal $1/f$ noise for healthy individuals and in the $1/f^2$ noise for pathological individuals. This condition is examined herein with the help of Fig. 1. The SOTC of Ref. \cite{korosh} yields $\mu < 2$. We examined the case of $\mu$ moving in the interval $2<\mu<3$ using  subordination to regular oscillatory motion, a phenomenological way of combining crucial events and periodicity. We believe that SOTC can be extended to this condition, and  hope that future work may realize this important goal. We see that for $f \rightarrow 0$, the IPL spectrum $S_{p}(f) \propto 1/f^{\beta}$, with $\beta = 3 - \mu$.  The ideal condition of $1/f$ noise is realized when $\mu = 2$.  The transition from $1/f$ noise to white  noise is  realized by increasing $\mu$ from the ideal value $\mu = 2$ to the value $\mu = 3$ and beyond.  In the presence of periodicity, though, the $1/f$ noise region can also be affected by moving the periodicity peak from the right to the left, in such a way as to make the $1/f^2$ noise become the predominant contribution to the spectrum in accordance with the experimental observation of Ref. \cite{physiology2}.

\emph{Monotasking versus Multitasking}: The theoretical perspective adopted herein affords an efficient way to approach this problem, while suggesting  an interesting approach to cognition.  In a recent paper Gershenson \cite{buddhism1} addresses the important issue of the connection between cognition and information and writes: ``Just like Buddhist philosophy, information theory and current cognitive science are pointing towards a worldview not centered on objective phenomena
(studied traditionally by physics), but centered on information, which can represent object, subject, and action within the same formalism".  The readers can find another attempt at making progress on the cognition issue in the paper \cite{buddhism2}. These authors rest on Buddhism to create a bridge between consciousness as a phenomenon in the operational architectonics of brain organization and quantum mechanics. The earlier work of these authors led to the discovery of rapid transition processes (RTP) that have been studied  in \cite{menicucci}, and found to be crucial events with $\mu >2$, but very close to $\mu = 2$.  For this reason, we are convinced that the theoretical approach adopted herein 
may help to build the bridge between West and East that the authors of \cite{buddhism1} and \cite{buddhism2} are trying to establish. It is very encouraging to notice 
that Tuladhar \emph{et al}  \cite{tuladhar} have recently found that meditation has the surprising effect of enhancing heartbeat coherence generating effects qualitatively similar to those illustrated in Fig. 1 of this paper, thereby leading us to interpret meditation as a mental process reducing variety to help the realization of specific tasks.

\emph{Homeodynamics}: The important results established herein are based on the action of crucial events, which  are a manifestation of temporal complexity, thereby explaining  why we  replace Ashby's homeostasis with homeo-dynamics. 

Finally, we conclude this paper stressing that the surprisingly accurate synchronization of the walking together process ought not to be confused with either chaos synchronization or resonance. In fact, chaos synchronization requires finite Lyapunov coefficients and resonance requires frequency tuning.  Complex systems with $\mu$ very close to the ideal condition $\mu = 2$, where the traditional Lyapunov coefficient vanishes,  have the affect of transferring their temporal complexity to systems with higher values of $\mu$. The numerical results show, that, although  communication through frequencies still exists (bottom panel of Fig. 4), the action of crucial events is more important for the transfer of intelligence. Our theoretical approach is based on the essential role of crucial events. The crucial events with $\mu$ becoming closer to $\mu =2$ are generators of multifractality, as pointed out in Refs.\cite {Multifractal} and \cite{bohara}. Thus, our prediction  that the walker with $\mu$ close to $2$ attracts the $\mu$ of the walker close to the Gaussian region $\mu = 3$ can be interpreted as a transmission of multifractality from the healthy to the sick walker in a surprising agreement with the recent experimental result of \cite{surprising}.

\emph{Acknowledgments} The authors thanks Dr.  Herbert Jelineck drawing our attention to Ref. \cite{physiology2}, PG thanks ARO for financial support of this work through grant W911NF1901.

\end{document}